\documentclass[preprint,showpacs,preprintnumbers,amsmath,amssymb]{revtex4}

\usepackage{graphicx}% Include figure files
\usepackage{dcolumn}% Align table columns on decimal point
\usepackage{bm}% bold math

\begin{document}

%\preprint{APS/123-QED}
\title{A new scheme to calculate the exchange tensor and its application to
  diluted magnetic semiconductors}

\author{H.~Ebert and
S.~Mankovsky}
\address{Universit\"at ~ M\"unchen, ~ Dept. ~ Chemie und
  Biochemie/Phys. Chemie, ~ Butenandtstr. ~ 5$-$13, ~ 
D$-$81377 ~ M\"unchen, ~ Germany}
\date{\today}% It is always \today, today,

\begin{abstract}
A new scheme to calculate the exchange tensor
$\underline{\underline{J}}_{ij}$ describing in a phenomenological way
the anisotropic exchange coupling of two moments in a magnetically
ordered system is presented. The ab-initio approach is based on spin-polarised
relativistic multiple-scattering theory within the framework of
spin-density functional theory. The scheme is applied to ferromagnetic
CrTe as well as the diluted magnetic semiconductor (DMS) system
Ga$_{1-x}$Mn$_{x}$As. In the later case the results show that there is a
noticeable anisotropy in the exchange coupling present, although not as
pronounced as suggested in recent theoretical investigations.
\end{abstract}

\pacs{Valid PACS appear here}% PACS, the Physics and Astronomy
%Use showkeys class option if keyword
\maketitle

% main text
%%%%%%%%%\section{}
\label{}

% \appendix

% \section{}
% \label{}

\section{Introduction}
The mapping of the energy of a magnetic solid calculated from first
principles for different magnetic configurations onto a Heisenberg
Hamiltonian is nowadays a widely used concept that allows a number of
interesting subsequent investigations. An example for this is the
determination of the Curie temperature of a ferromagnet by means of
Monte Carlo simulations using the calculated exchange coupling
parameters $J_{ij}$ as input ($i$ and $j$ are indices labeling the individual
atomic sites) \cite{BKBB03}.
Apart from fitting the $J_{ij}$'s to the total
energies obtained for different magnetic configurations one can use the
energies 
of spin spirals as basis for such a mapping \cite{HEPO98}. As an
alternative one 
may use perturbation theory, that allow to calculate the $J_{ij}$'s
directly. In fact the expression derived by Lichtenstein et al. \cite{LKAG87} within
the frame work of non-relativistic multiple scattering theory is now
successfully used for a wide range of materials \cite{PKT+01,PKT+00,SDK06}.

 Initiated among others by investigations on the magnetic ground state
 configuration of  nano-scale systems, there is strongly growing
 interest in the interplay of exchange coupling and spin-orbit
 coupling \cite{FBV+08}. Besides the magnetic anisotropy energy, the spin-orbit
 coupling gives rise to an anisotropic exchange coupling. Using again
 the above mentioned concept the isotropic exchange constants $J_{ij}$ have
 to be replaced by a corresponding exchange coupling tensor
 $\underline{\underline{J}}_{ij}$. By 
 generalising the approach of Lichtenstein et al. to a fully
 relativistic formulation, Udvardi et al. \cite{USPW03} could derive
 corresponding 
 expressions for the elements of $\underline{\underline{J}}_{ij}$. Corresponding applications to thin
 films as well as to finite deposited clusters can be found in the
 literature \cite{USPW03,ALU+08}. A disadvantage of the expressions worked out by Udvardi
 et al. \cite{USPW03} is that one has to use various magnetic
 configurations as a reference state to determine all elements of
 $\underline{\underline{J}}_{ij}$. In the following we 
 present an alternative approach that can be derived in a rather
 transparent way and does not have that problem.  
As it is demonstrated both approaches give nevertheless results that are
quite close to each other. As an application of our new scheme we
present results for the exchange tensor $\underline{\underline{J}}_{ij}$
in ferromagnetic CrTe and in the diluted magnetic semiconductor (DMS) system
Ga$_{1-x}$Mn$_{x}$As. The later
will be discussed in relation to recent work of Timm and MacDonald
\cite{TM05} who used a tight-binding description of the system in
contrast to the ab-initio approach employed here that is based on local
spin density approximation (LSDA).

\section{Theoretical approach}
Starting point of our derivation for $\underline{\underline{J}}_{ij}$ is
the expression for the 
change in energy $\Delta E_{ij}$ of a system upon a perturbation taking place
at sites $i$ and $j$. An expression for $\Delta E_{ij}$ was worked out
by several 
authors \cite{LKAG87,OTH83} within the framework of
multiple scattering 
theory and making use of Lloyd's formula.  
The derivation of this expression can straight forwardly be applied when
working in the framework of spin-polarised relativistic multiple
scattering or KKR \cite{Ebe00} formalism. Adopting the
convention for the corresponding electronic Green's function as used by
Dederichs and coworkers \cite{DDZ92} its off-site part is given by: 

\begin{eqnarray}
G(\vec{r}_i,\vec{r}_j,E) &=& -ip \sum_{\Lambda\Lambda'}
R^{i}_{\Lambda}(\vec{r}_i,E) G^{ij}_{\Lambda\Lambda'}(E)
R^{j\times}_{\Lambda'}(\vec{r}_j,E) \; , 
\end{eqnarray}
where $G^{ij}_{\Lambda\Lambda'}(E)$ is the so-called structural Green's
function, $R^{i}_{\Lambda}$ is a 
regular solution to the single-site Dirac equation labeled by the
combined quantum numbers $\Lambda$ ($\Lambda = (\kappa,\mu)$), with
$\kappa$ and $\mu$  being the spin-orbit and magnetic quantum numbers
\cite{Ros61} and $p$ is the electron momentum. The energy change $\Delta
E_{ij}$ can then be written as \cite{LKAG87,OTH83}  

\begin{eqnarray}
\Delta E_{ij} & = & -\frac{1}{\pi} \Im \int dE \mathrm{Trace} \Delta
\underline{t}^i  \underline{G}^{ij} \Delta\underline{t}^j
\underline{G}^{ji} \;,
\end{eqnarray}
where $\Delta \underline{t}^{i}$ is the change of the single-site
t-matrix due to the perturbation $\Delta V^{i}(\vec{r})$ at site $i$ and
the underline denotes matrices with 
respect to the quantum numbers $\Lambda$. To first order in $\Delta
V^{i}(\vec{r})$  the change $\Delta \underline{t}^{i}$ is given by  

\begin{eqnarray}
\Delta t_{\Lambda'\Lambda}^i & = & \int d^3r
R^{i\times}_{\Lambda'}(\vec{r})\Delta V(\vec{r})R^{i}_\Lambda(\vec{r}) = \Delta
V_{\Lambda'\Lambda}^{(R)i}  \; .  
\label{Eq_3}
\end{eqnarray}

Using instead the convention for the Green's function as used by Gy\"orffy
and coworkers \cite{Wei90} one may express $\Delta E_{ij}$ in terms of the
scattering path operator $\tau^{ij}_{\Lambda'\Lambda}(E)$

\begin{eqnarray}
\Delta E_{ij} & = & -\frac{1}{\pi} \Im \int dE \mathrm{Trace} 
\Delta \underline{V}^{(Z)i} \underline{\tau}^{ij}  \Delta
\underline{V}^{(Z)j} \underline{\tau}^{ji} \; ,
\end{eqnarray}
where use have been made of the relation  $\underline{G}^{ij} =
(\underline{t}^i)^{-1}  \underline{\tau}^{ij} (\underline{t}^j)^{-1}$
for $i \neq j$ 
and the matrix elements $\Delta V^{(Z)i}_{\Lambda\Lambda'}$
 are to be evaluated using the alternative set of regular solutions  $Z_{\Lambda}^i$ to the single-site Dirac equation \cite{Ebe00,Wei90}. 

Changing the orientation of the spin magnetic moment $\vec{m}_i$ within an
atomic cell $i$ and adopting the rigid spin approximation (RSA) \cite{AKH+96}
implies a corresponding change of the spin-dependent potential $\beta \vec{\sigma}\vec{B}(\vec{r})$  by:

\begin{eqnarray}
\Delta V(\vec{r}) & = & V_{\hat{n}}(\vec{r}) -  V_{\hat{n}_0}(\vec{r}) 
  =  \beta \vec{\sigma}(\hat{n} - \hat{n}_0)B(\vec{r}) \; ,
\label{Eq_5}
\end{eqnarray}
where $\beta$ is one of the standard Dirac matrices and $\vec{\sigma}$
is the vector of $4\times4$-spin matrices \cite{AKH+96}. In writing
Eq. (\ref{Eq_5}) a collinear spin magnetisation within the cell has been
assumed together with a change of its orientation from $\hat{n}_0$ to 
$\hat{n}$. Accordingly, $B(\vec{r})$ corresponds to the difference
of the spin-projected potential functions ${B}(\vec{r}) =
\frac{1}{2}(V^{\uparrow}(\vec{r}) - V^{\downarrow}(\vec{r}))$
\cite{Ebe00}.  This leads for the matrix elements $\Delta 
V^{(Z)i}_{\Lambda\Lambda'}$ to:

\begin{eqnarray}
\Delta V^{(Z)i}_{\Lambda\Lambda'} & = &
 \sum_{\alpha = x,y,z}\Delta
V^{(Z)i\alpha}_{\Lambda\Lambda'} \Delta \alpha 
\end{eqnarray}

with

\begin{eqnarray}
\Delta V^{(Z)i\alpha}_{\Lambda\Lambda'} & = & \int d^3r
Z^{\times}_{\Lambda}(\vec{r})\beta \sigma_{\alpha}B(r)
Z_{\Lambda'}(\vec{r}) \; .
\end{eqnarray}

Comparison with the generalised Heisenberg Hamiltonian

\begin{eqnarray}
H_{ex} & = & - \frac{1}{2}\sum_{i,j} \hat{e}_i  \underline{\underline{J}}_{ij}
\hat{e}_j 
\label{Eq_8}
\end{eqnarray}

with $\hat{e}_{i(j)}$ the orientation of the spin magnetic moment at site
$i (j)$ allows one to write for the elements of the exchange coupling
tensor $\underline{\underline{J}}_{ij}$

\begin{eqnarray}
 J_{ij}^{\alpha_i \alpha_j}  &=&  -\frac{1}{\pi} \Im \int dE \mathrm{Trace} 
\Delta \underline{V}^{(Z)\alpha_i} \underline{\tau}^{ij}
\Delta \underline{V}^{(Z)\alpha_j} \underline{\tau}^{ji} \; .
\end{eqnarray}

The scheme outlined above has been implemented using the spin-polarised
relativistic (SPR) version of multiple scattering theory
\cite{Ebe00,Wei90}. All 
calculations have been done within the framework of the local spin
density approximation (LSDA) to spin density functional theory
\cite{VWN80}. To represent the results for the exchange tensor
$\underline{\underline{J}}_{ij}$ we use the conventional decomposition
of the corresponding Heisenberg Hamiltonian in Eq. (\ref{Eq_8}) \cite{USPW03}:

\begin{eqnarray*}
 H_{ex} &=&
 -\frac{1}{2} \sum_{ij}
 \hat{e}_iJ_{ij}\hat{e}_j 
 -\frac{1}{2} \sum_{ij}
 \hat{e}_i\underline{\underline{J}}^S_{ij}\hat{e}_j \\
&& - \frac{1}{2} \sum_{ij} \vec{D}_{ij}[\hat{e}_i\times \hat{e}_j] \; .
\end{eqnarray*}
Here $J_{ij}$ is the isotropic exchange coupling constant,
$\underline{\underline{J}}^S_{ij}$ is the traceless symmetric part of
$\underline{\underline{J}}_{ij}$ and the antisymmetric part is
represented by the Dzyaloshinski-Moriya (DM) vector $\vec{D}_{ij}$. It
should be emphasized that {\rm isotropic} in the context of $J_{ij}$ refers
to spin-space and does not imply that there in no anisotropy in real
space, i. e. $J_{ij}$ will in general not only depend on the distance
$|\vec{R}_{ij}|$ between two sites but also on the orientation
$\hat{R}_{ij}$ of the distance vector.

%FFFFFFFFFFFFFFFFFFFFFFFFFFFFFFFFFFFFFFFFFFFFFFFFFF
\begin{figure}
\begin{center}
\includegraphics[width=6cm,clip]{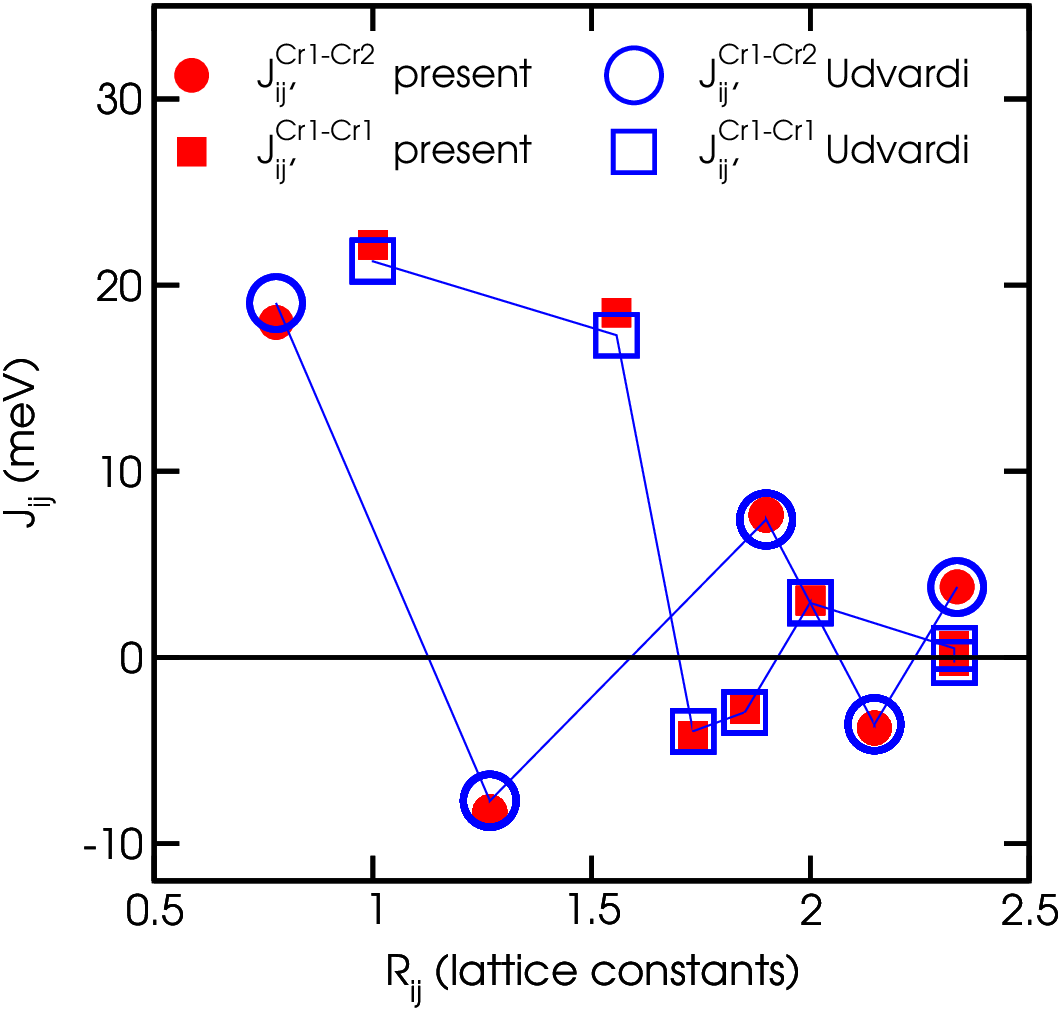}
\caption{\label{fig_1}Isotropic exchange interaction parameters $J_{ij}$
  between Cr atoms at sites $i$ and $j$ in ferromagnetic CrTe as a
  function of the  inter-atomic distance $R_{ij}$. The results based on
  the present 
  approach (full symbols) are compared to results obtained using the
  approach of Udvardi et al. \cite{USPW03} (open symbols). Circles and
squares represent the coupling of a Cr atom in layer 1 (Cr1) to another
Cr atom in layer 1 (Cr1) or layer 2 (Cr2), respectively.}  
\end{center}
\end{figure}
%FFFFFFFFFFFFFFFFFFFFFFFFFFFFFFFFFFFFFFFFFFFFFFFFFF

%FFFFFFFFFFFFFFFFFFFFFFFFFFFFFFFFFFFFFFFFFFFFFFFFFF
\begin{figure}
\begin{center}
\includegraphics[width=6cm,clip]{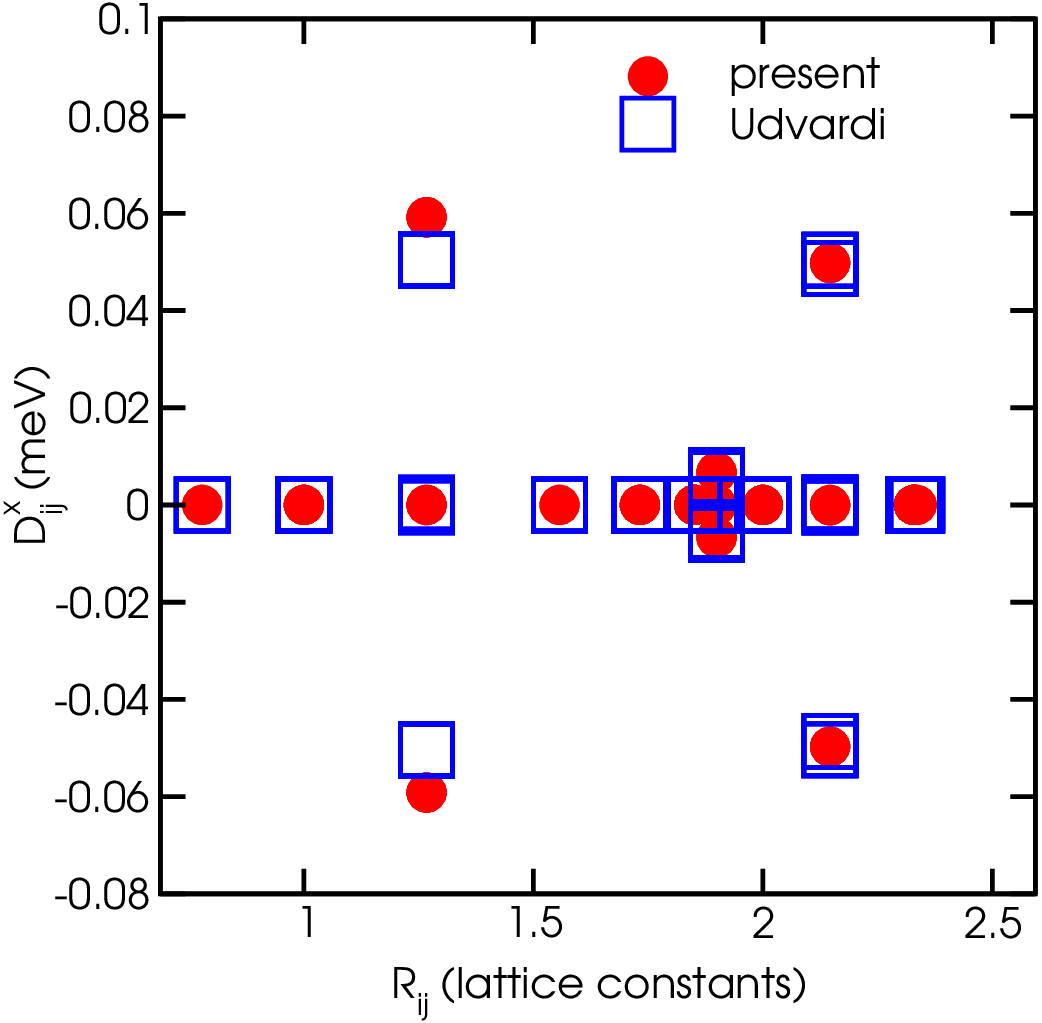}\\
\includegraphics[width=6cm,clip]{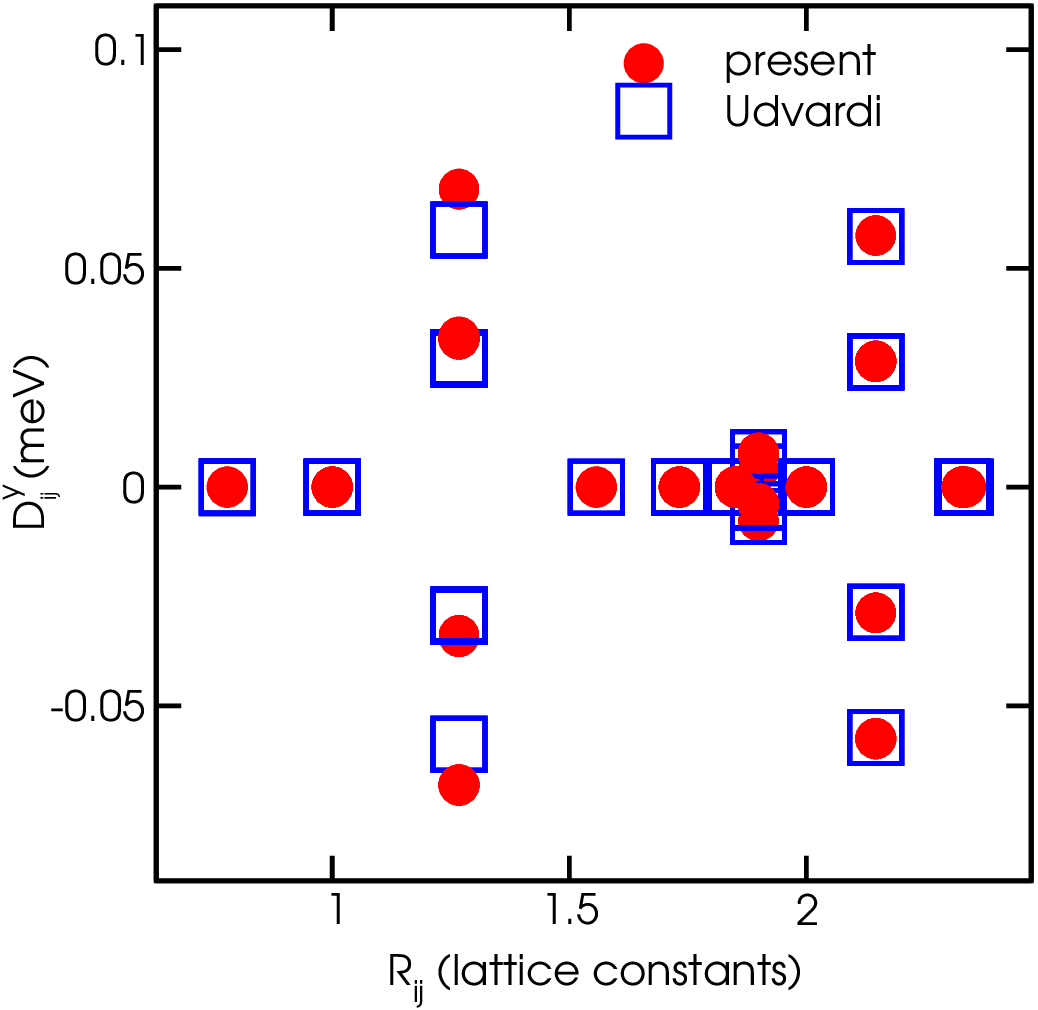}\\
\includegraphics[width=6cm,clip]{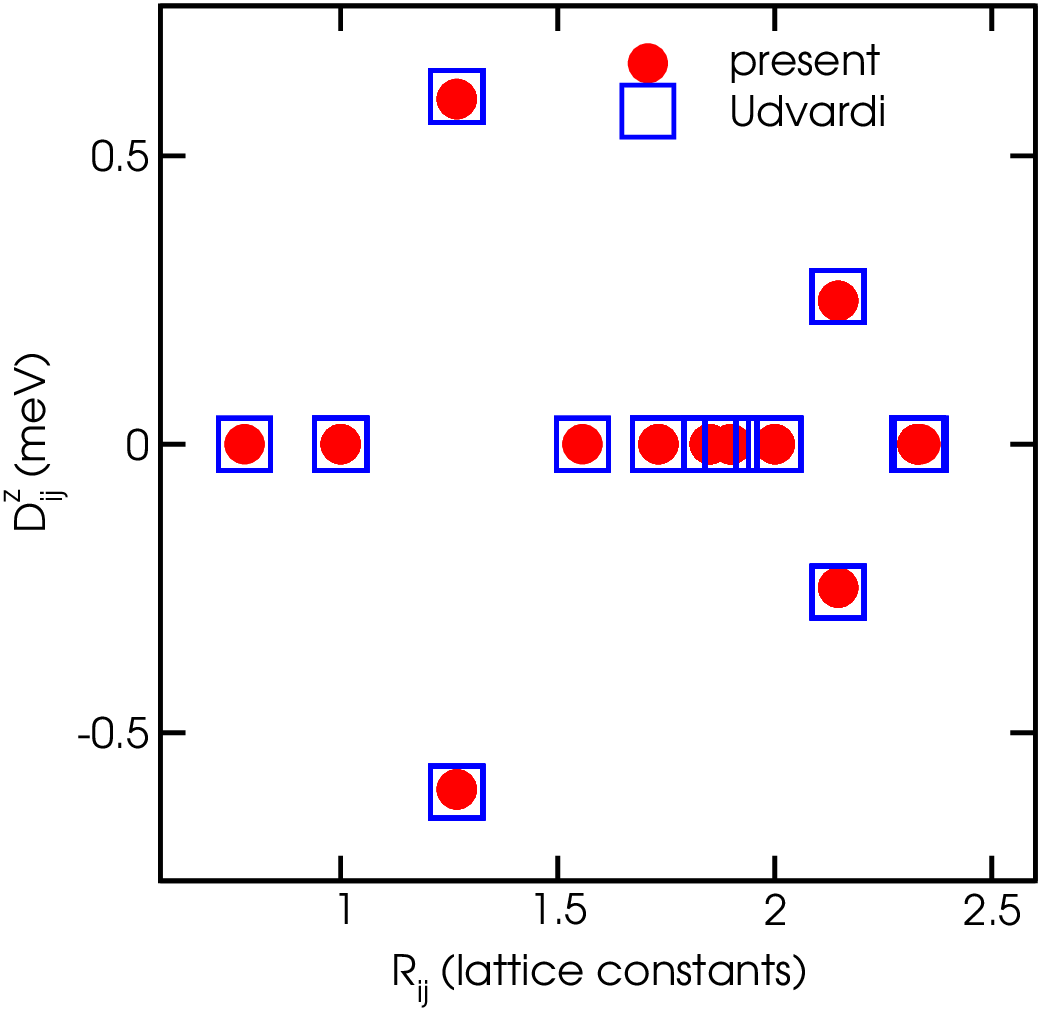}
\caption{\label{fig_2} Components of the Dzyaloshinski-Moriya interaction
 vector $\vec{D}_{ij}$ (from top to bottom: $x, y, z$) between Cr 
  atoms at sites $i$ and $j$ in ferromagnetic CrTe as a function of the
  inter-atomic distance $R_{ij}$. The results based on the present
  approach (full circles) are compared to results obtained using the
  approach of Udvardi et al. \cite{USPW03} (open squares). 
  }
\end{center}
\end{figure}
%FFFFFFFFFFFFFFFFFFFFFFFFFFFFFFFFFFFFFFFFFFFFFFFFFF

\section{Results and discussion}
To demonstrate the application of our approach we present in Figs.
\ref{fig_1} and  \ref{fig_2} results for the coupling parameters
$J_{ij}$ and $\vec{D}_{ij}$ of the two inequivalent Cr atoms in ferromagnetic CrTe
having the NiAs structure. The isotropic parameters $J_{ij}$ shown in
Fig. \ref{fig_1} reflect dominating ferromagnetic coupling that is quite
far reaching, i.e. slowly decaying. 
As one notes the isotropic exchange coupling between a central Cr atom
in layer 1 (denoted Cr1) to another Cr atom in layer 1 and 2 (denoted
Cr1 and Cr2 and represented by squares and circles, respectively, in
Fig. \ref{fig_1}) is in the same order of magnitude. This means there is
no remarkable spatial anisotropy imposed by the layered structure of the
system for the isotropic coupling constant $J_{ij}$.
The anisotropy of the exchange
coupling is represented by  $\underline{\underline{J}}^S_{ij}$ as well
as by $\vec{D}_{ij}$. As $\underline{\underline{J}}^S_{ij}$ turns out
to be quite small we show in Fig. \ref{fig_2} only the three components
of the DM vector $\vec{D}_{ij}$. Many of the DM vector components are
zero due to symmetry restrictions \cite{Mor60}. 
In particular one finds a non-zero DM vector $\vec{D}_{ij}$ only if the sites $i$ and $j$ belong to different sub-lattices Cr1 and Cr2.
Due to the NiAs
structure of the 
system the non-vanishing $x$- and $y$-components of the vector are of the
same order of magnitude  while the $z$-component is one order of magnitude
larger. The different behaviour of the $x$-, $y$- and $z$-components reflects
obviously to some extent the quasi-layered structure of the system
(hexagonal Cr layers with Te-layers in between). However, the
anisotropic exchange coupling is still about two orders of magnitude
smaller than the isotropic one.

%FFFFFFFFFFFFFFFFFFFFFFFFFFFFFFFFFFFFFFFFFFFFFFFFFF
\begin{figure}
\begin{center}
\includegraphics[width=6cm,clip]{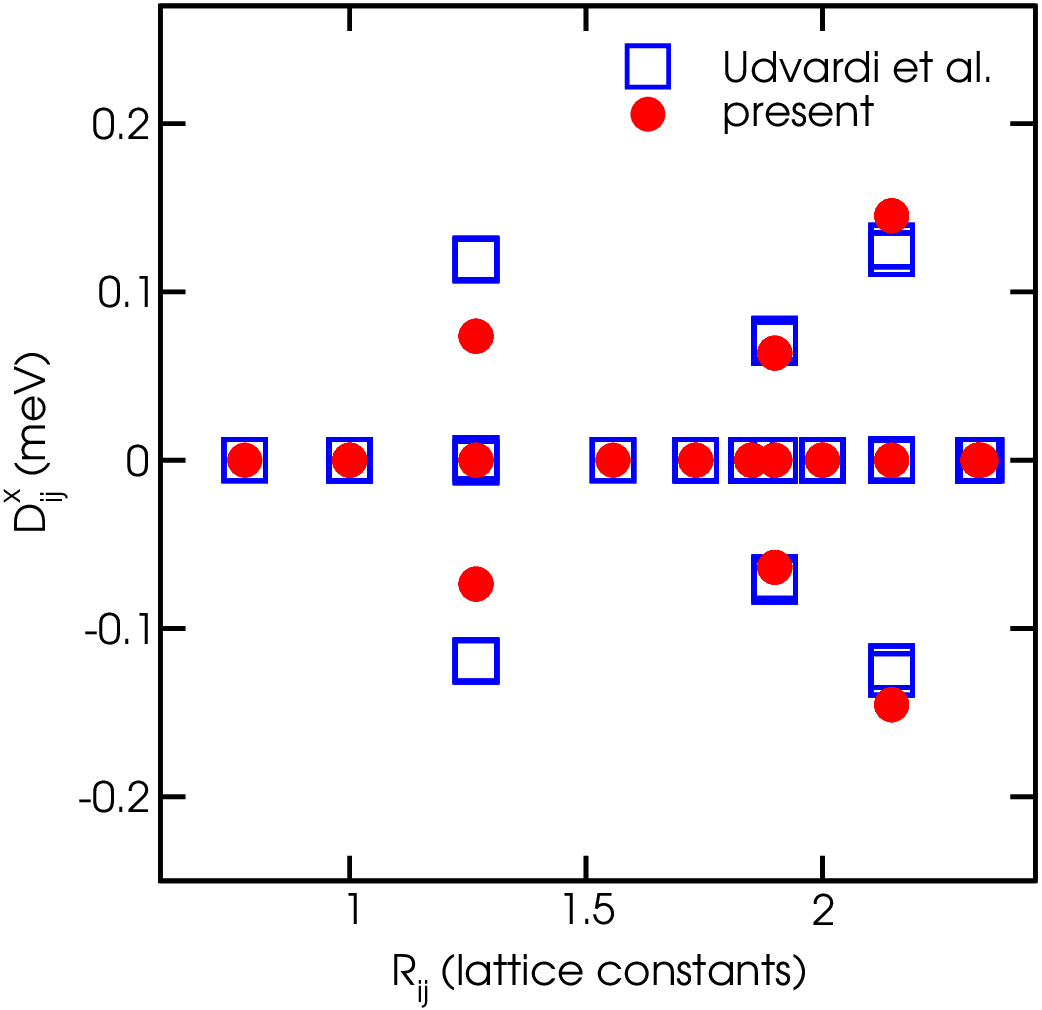}
\caption{\label{fig_2a} The same as in Fig. \ref{fig_2} top but with
  spin-orbit coupling scaled by factor 2.} 
\end{center}
\end{figure}
%FFFFFFFFFFFFFFFFFFFFFFFFFFFFFFFFFFFFFFFFFFFFFFFFFF

%FFFFFFFFFFFFFFFFFFFFFFFFFFFFFFFFFFFFFFFFFFFFFFFFFF
\begin{figure*}
\begin{center}
\includegraphics[width=6cm,clip]{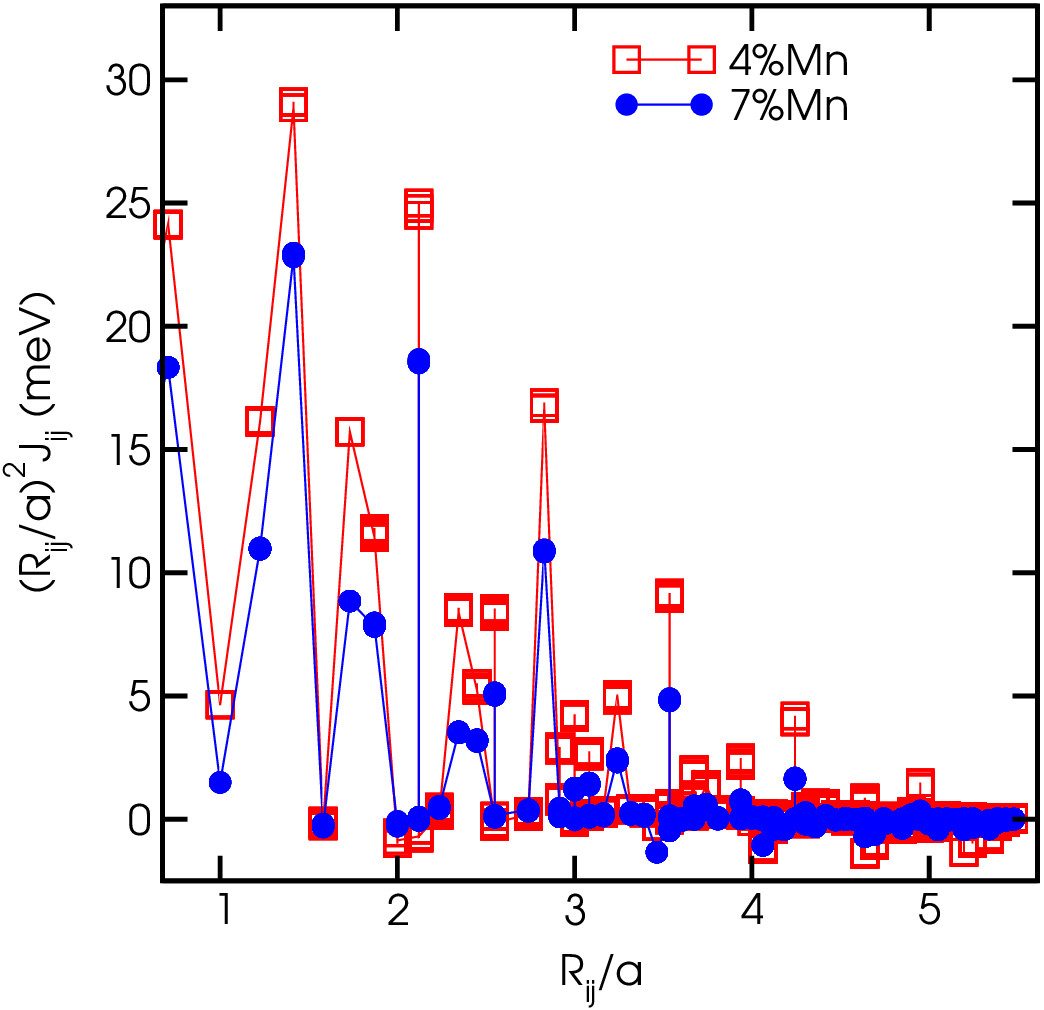}
\caption{\label{fig_3} Isotropic exchange interaction $J_{ij}$ between Mn
  atoms at sites $i$ and $j$ in Ga$_{1-x}$Mn$_{x}$As, scaled by the factor 
  $(R_{ij}/a)^2$, as a function of the 
  inter-atomic distance $R_{ij}$. Results are given for $7$ at.$\%$
  Mn (full circles) and for $4$ at.$\%$ Mn (open squares).} 
\end{center}
\end{figure*}
%FFFFFFFFFFFFFFFFFFFFFFFFFFFFFFFFFFFFFFFFFFFFFFFFFF

%FFFFFFFFFFFFFFFFFFFFFFFFFFFFFFFFFFFFFFFFFFFFFFFFFF
\begin{figure}[r]
\begin{center}
\includegraphics[width=6cm,clip]{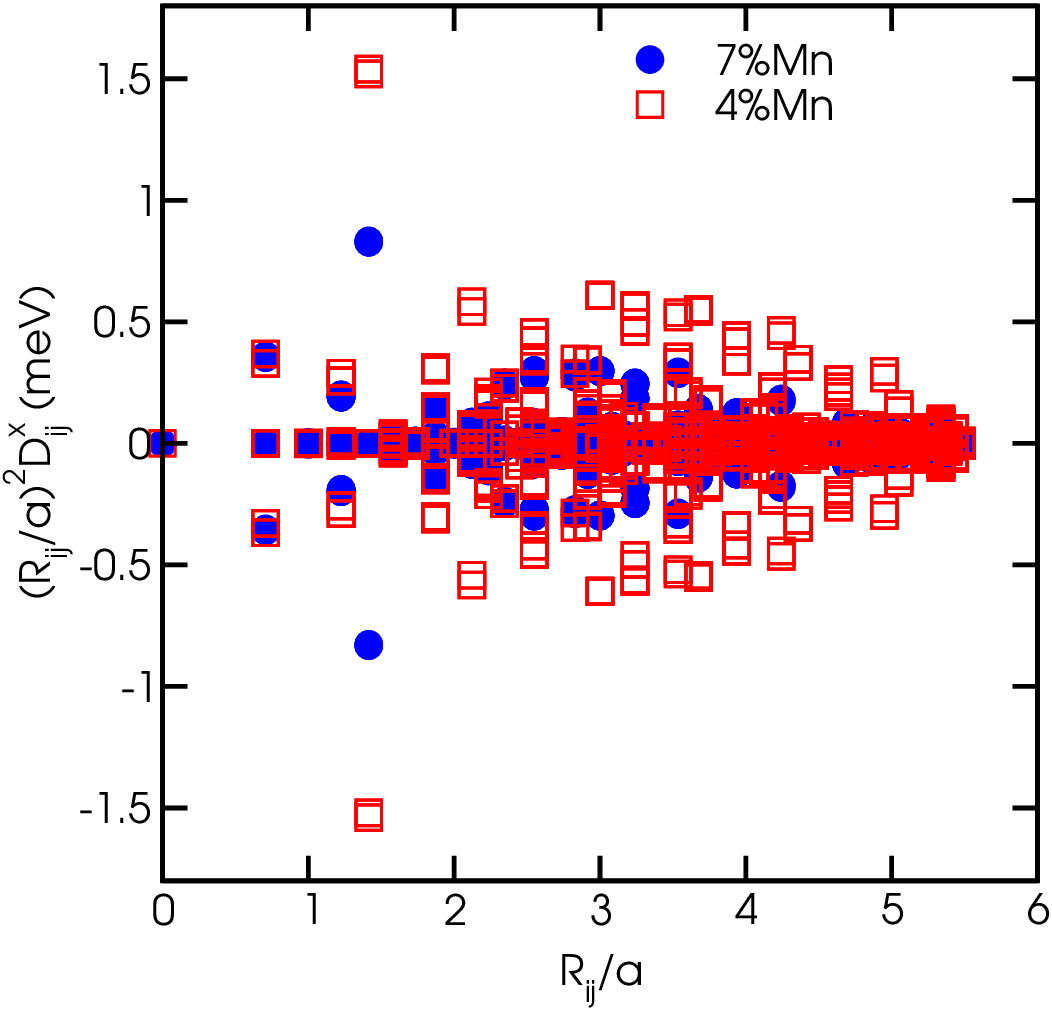}\\
\includegraphics[width=6cm,clip]{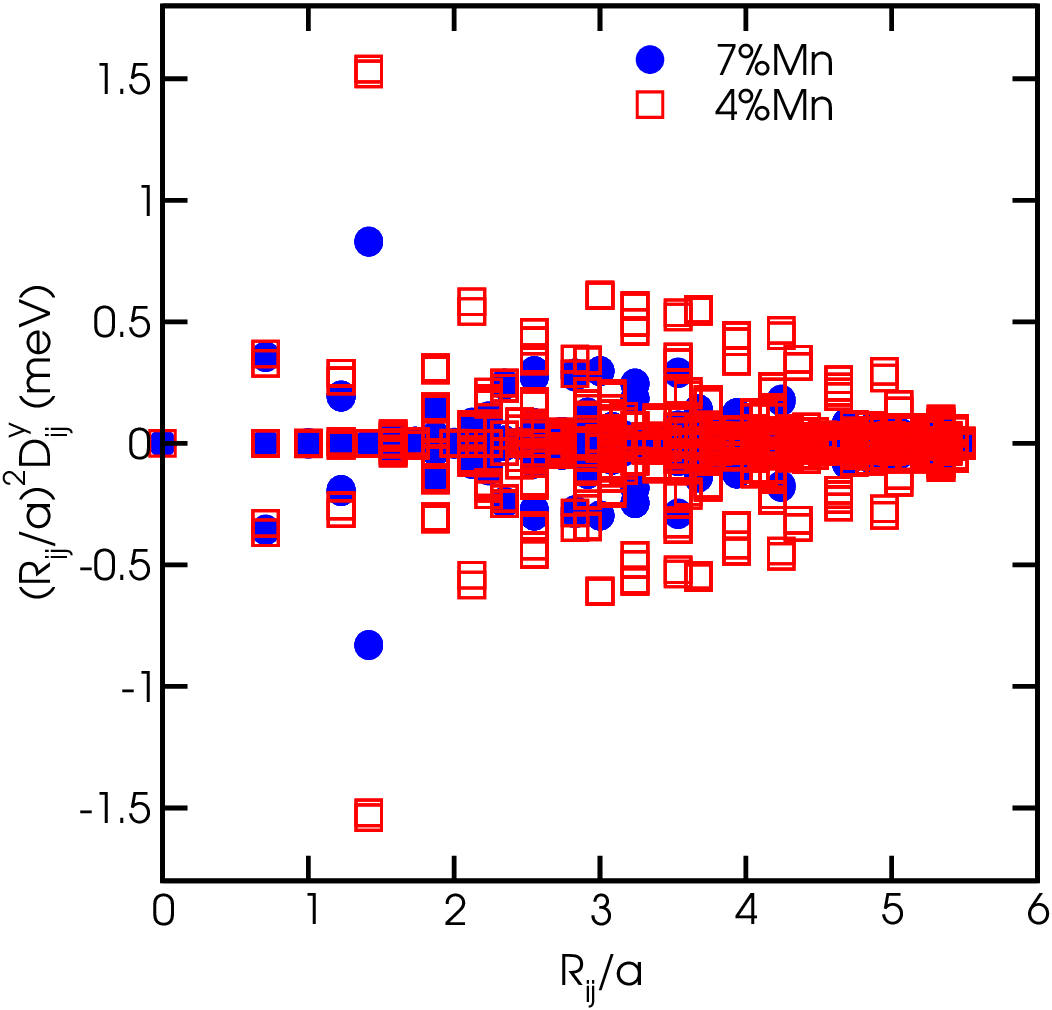}\\
\includegraphics[width=6cm,clip]{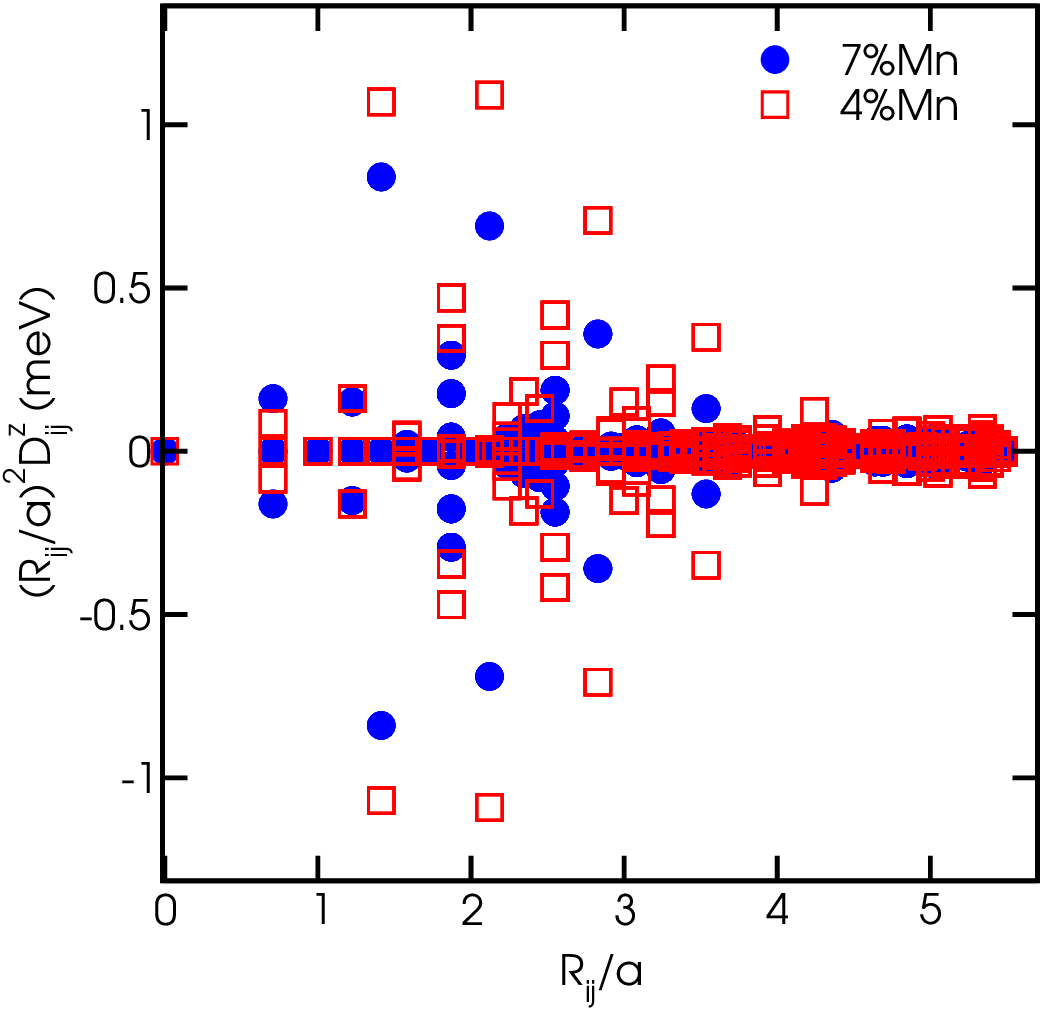}
\caption{\label{fig_4} Components of Dzyaloshinski-Moriya interaction vector
  $\vec{D}_{ij}$ (from top to bottom: $x, y, z$) between Mn atoms at sites $i$ and $j$ in
  Ga$_{1-x}$Mn$_{x}$As, scaled by the factor $(R_{ij}/a)^2$, as a
  function of the inter-atomic distance $R_{ij}$. Results are
  given for $7$ at.$\%$ Mn (full circles) and for $4$ at.$\%$ Mn (open
  squares).} 
\end{center}
\end{figure}
%FFFFFFFFFFFFFFFFFFFFFFFFFFFFFFFFFFFFFFFFFFFFFFFFFF

As the comparison of the results for $J_{ij}$ and $\vec{D}_{ij}$
obtained using the approach presented above and that of Udvardi et
al. \cite{USPW03}, respectively, in Figs. \ref{fig_1} and \ref{fig_2}
demonstrates, both schemes give very similar results. This also holds
for other systems studied so far with most pronounced differences
occurring for the DM vector. It should be stressed, however, that the above
scheme allows to determine $\underline{\underline{J}}_{ij}$ with respect
to one common reference state; i.e.\ there is no need to use various reference
states to get all tensor elements. This ensures that the results for the
various elements are always consistent even when the choice of the
reference state influence the result, e.g. when the RSA is not fully
justified. 

To demonstrate that the DM interaction is indeed induced by spin-orbit
coupling (SOC) we performed model calculations with the strength of SOC
artificially increased by a factor of 2. While the isotropic exchange
coupling constants $J_{ij}$ hardly changed, the DM vector components
increased essentially by the same factor. This can be seen by comparing
$D^x_{ij}$ given  in Fig. \ref{fig_2a} with the results in
Fig. \ref{fig_2} (top). Fig. \ref{fig_2a} also shows that for this
specific situation results based on the two approaches considered may
differ in an appreciable way.

%FFFFFFFFFFFFFFFFFFFFFFFFFFFFFFFFFFFFFFFFFFFFFFFFFF
\begin{figure}
\begin{center}
\includegraphics[width=6cm,clip]{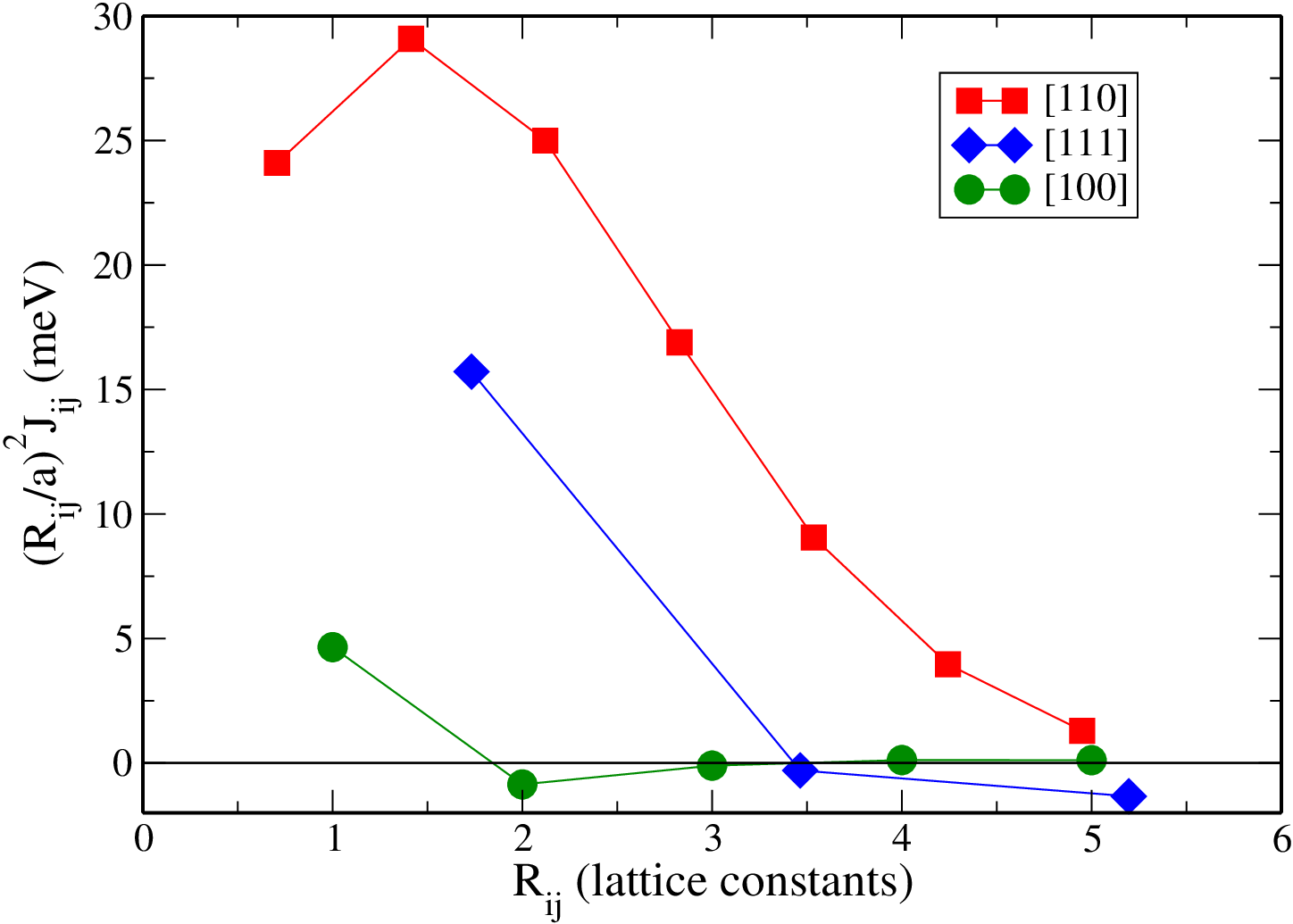}
\caption{\label{fig_6}  Scaled
  exchange interactions $(R_{ij}/a)^2J_{ij}$ in 
  ferromagetic Ga$_{1-x}$Mn$_{x}$As, $x=0.04$, as a function of the
  inter-atomic distance $R_{ij}$ along different directions.
 }
\end{center}
\end{figure}
%FFFFFFFFFFFFFFFFFFFFFFFFFFFFFFFFFFFFFFFFFFFFFFFFFF

%FFFFFFFFFFFFFFFFFFFFFFFFFFFFFFFFFFFFFFFFFFFFFFFFFF
\begin{figure}
\begin{center}
\includegraphics[width=6cm,clip]{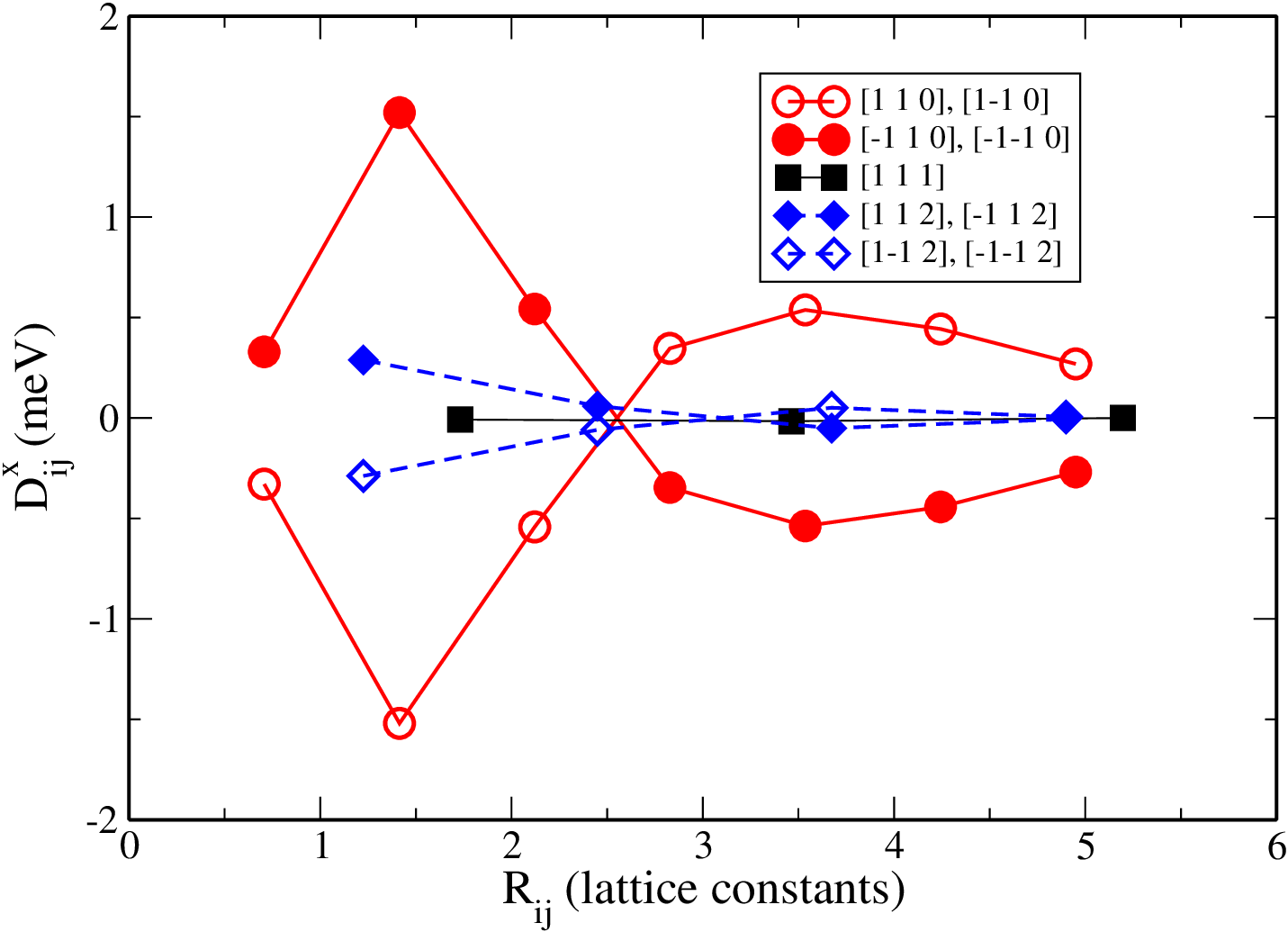}
\caption{\label{fig_8}  Components of Dzyaloshinski-Moriya interaction
  $\vec{D}_{ij}$ between Mn atoms at sites $i$ and $j$ in
  Ga$_{1-x}$Mn$_{x}$As, $x=0.04$, scaled by factor $(R_{ij}/a)^2$, 
as a function of the inter-atomic distance $R_{ij}$ along different
directions. 
 }
\end{center}
\end{figure}
%FFFFFFFFFFFFFFFFFFFFFFFFFFFFFFFFFFFFFFFFFFFFFFFFFF

Figs.\ \ref{fig_3} and  \ref{fig_4} show the results of an application of
our approach to the diluted magnetic semiconductor system
Ga$_{1-x}$Mn$_{x}$As. The isotropic exchange coupling (Fig. \ref{fig_3})
and also its 
concentration dependence agrees quite well with the results of other authors
\cite{SDK06}, indicating in particular that the spin-orbit coupling
accounted within the present work affects the isotropic exchange
coupling only slightly. 
As mentioned above and as was noted by other authors \cite{SDK06,KTD+04}
there is a directional dependency for  $J_{ij}$. This is demonstrated by
Fig. \ref{fig_6} where results for the concentration $x=0.04$ are given
for $\hat{R}_{ij}$ along [001], [110], and [111] separately. As for the
magnitude of $J_{ij}$, this spatial anisotropy of   $J_{ij}$ is only
slightly influenced by inclusion of SOC.
In contrast to this SOC is ultimately responsible
for the anisotropy in the exchange coupling represented by the DM vector
shown in Fig. \ref{fig_4}. In contrast to the CrTe system considered
above the three components of the DM vectors are of the same order of
magnitude as a consequence of the zincblende structure of
Ga$_{1-x}$Mn$_{x}$As. 

Although, it seems not possible to give a simple
scaling behaviour of the magnitude of the exchange coupling parameters
with respect to the inter-atomic distance $R_{ij}$, one notes that the
components of $\vec{D}_{ij}$ decay less rapidly as $J_{ij}$ with
increasing $R_{ij}$. This behaviour was also found for other systems and
is in line with the findings of Timm and MacDonald \cite{TM05}. However, 
our results for the isotropic as well as anisotropic
exchange coupling constants differ quite appreciably from those obtained
recently in a more phenomenological way by these authors. In
particular the tensor elements representing anisotropic exchange are
found to be around one order of magnitude smaller than given in the
previous work. The anisotropy of the DM interaction is demonstrated in
Fig. \ref{fig_8} for the component $D^x_{ij}$. As one notes $D^x_{ij}$
depends quite strongly in the direction $R_{ij}$. In particular one
finds $D^x_{ij}$ to be zero e.g. for the [111] direction due to
symmetry. Also because of the symmetry of the system one finds for each
direction a symmetry-related one for which the sign of $D^x_{ij}$ is
reversed.

The presence of a non-collinear ferromagnetic structure in
Ga$_{1-x}$Mn$_x$As is 
assumed to be partially responsible for the missing of remanent
magnetisation observed experimentally in annealed samples \cite{PKC+01,
PKM+02, PKW+03}. In particular, the remanent magnetisation in
 this DMS system can be noticably increased in the presence
of a rather small magnetic field. Such a behaviour could indeed be
explained by the presence of non-collinear magnetism in the system \cite{PKW+03}.   
The anisotropy of the exchange coupling in Ga$_{1-x}$Mn$_x$As 
was studied theoretically by various authors \cite{SM02,ZJ02,BG03,TM05} to
find whether it 
can be responsible for the formation of a non-collinear
ferromagnetic structure as a ground state in this DMS system. 
However, these investigations were based on phenomenological or
semi-phenomenological approaches and the results obtained 
are rather controversial. 
In contrast to this, the present approach allows us to evaluate the
elements of the exchange 
coupling tensor (in particular, its antisymmetric part 
representing the DM coupling) on the basis of ab-initio electronic
structure calculations. As was demonstrated above, this leads indeed to
a rather large  value for the DM coupling in Ga$_{1-x}$Mn$_x$As, which
is only about one order of magnitude smaller than for the 
isotropic exchange. As mentioned, this finding is in line with the
results of Timm and MacDonald \cite{TM05}. Obviously, the values for the
DM coupling term cannot be considered as negligibly small, and as a
consequence one cannot 
exclude a noticable non-collinear ferromagnetic order in the system. To
clarify this question corresponding Monte Carlo simulations based on the
calculated exchange tensor will be performed.

\section{Summary}

A new scheme to calculate the exchange coupling tensor
$\underline{\underline{J}}_{ij}$ has been presented that is based on
ab-initio electronic structure calculations using spin-polarized fully
relativistic 
multiple scattering theory and spin density functional
theory. Application to ferromagnetic CrTe as well as to other systems
demonstrates that the approach gives results for the exchange tensor
elements very similar to those obtained using the approach of Udvardi et
al. However, the new approach makes use of a unique reference state
ensuring the internal consistency of the tensor elements. Application to
the diluted magnetic semiconductor system Ga$_{1-x}$Mn$_{x}$As led to an
isotropic exchange in full accordance with previous non-relativistic
calculations that were also based on ab-initio electronic structure
calculations. The results obtained for the anisotropic exchange coupling
are in accordance with the data of Timm and MacDonald
concerning the variation with distance. However, numerically the
coupling constants obtained by the present ab-initio approach and the
more phenomenological scheme of these authors differ in an appreciable
way.

\section{Acknowledgements}

This work was supported by the Deutsche Forschungsgemeinschaft within
the Schwerpunktprogramm 1153, Schwerpunktprogramm 1136 as well as the
Sonderforschungsbereich SFB 689.

%
%%%%%%%%%%%%%%%%%%%%%%%%%%%%%%%%%%%%%%%%%%%%%%%%%%%%%%%%%%%%%%%%%%%%%%%%%%%
%                         REFERENCES
%%%%%%%%%%%%%%%%%%%%%%%%%%%%%%%%%%%%%%%%%%%%%%%%%%%%%%%%%%%%%%%%%%%%%%%%%%%
%\bibliography{akhelit,anisotropy}
%\bibliographystyle{ifac}
%\bibliographystyle{prsty}

\end{document}